\documentclass[a4paper,12pt]{article}
\usepackage{amsmath,latexsym,amssymb,graphicx,anysize}
\usepackage[square,numbers,comma,sort&compress]{natbib}
\marginsize{1.5cm}{1.5cm}{1cm}{1cm}
\newcommand{\be}{\begin{equation}}
\newcommand{\ee}{\end{equation}}
\newcommand{\bea}{\begin{eqnarray}}
\newcommand{\eea}{\end{eqnarray}}
\newcommand{\emailjps}{jpsinghmp@lycos.com}
\newcommand{\emailrb}{balir5@yahoo.co.in}
\newcommand{\emailps}{pratibhasingh274@gmail.com}

\title{Bianchi Type V Viscous Fluid Cosmological Models in Presence of Decaying Vacuum Energy}
\author{Raj Bali$^1$\footnote{e-mail: \emailrb},~~Pratibha Singh$^1$\footnote{e-mail: \emailps}~and~J. P. Singh$^2$\footnote{e-mail: \emailjps}\\
\small{$^1$Department of Mathematics, University of Rajasthan, Jaipur - 302004, India}\\
\small{$^2$Department of Mathematical Sciences, A. P. S. University, Rewa - 486003, India}}
\begin{document}
\date{}
\maketitle
\begin{abstract}
Bianchi type V viscous fluid cosmological model for barotropic fluid distribution with varying cosmological term $\Lambda$ is investigated. We have examined a cosmological scenario proposing a variation law for Hubble parameter $H$ in the background of homogeneous, anisotropic Bianchi type V space-time. The model isotropizes asymptotically and the presence of shear viscosity accelerates the isotropization. The model describes a unified expansion history of the universe indicating initial decelerating expansion and late time accelerating phase. Cosmological consequences of the model are also discussed.
\end{abstract}



\maketitle
\section{Introduction}
Bianchi type V cosmological models are interesting in the study because these models contain isotropic special cases and allow arbitrary small anisotropy levels at any instant of cosmic time. These models represent the open FRW (Friedmann-Robertson-Walker) model with $k = -1$, where k is the curvature of three  dimensional space.~\citet{rf301,rf302,rf303} have investigated Bianchi type V models in different contexts.~\citet{rf304} has  investigated  Bianchi  type V  cosmological  model  for  perfect  fluid distribution in general relativity.~\citet{rf305} have investigated LRS (Locally Rotationally Symmetric) Bianchi V space-time with shear and bulk viscosities.~\citet{rf306} have investigated Bianchi type V model with heat flow and viscosity.~\citet{rf307} studied Bianchi type V imperfect fluid cosmological models for barotropoic fluid distribution $p=\gamma\rho$, $p$ being isotropic pressure, $\rho$ the matter density and $0\leq\gamma\leq 1$.~\citet{rf308,rf309,rf310} have investigated Bianchi type V bulk viscous fluid cosmological models in general relativity in different contexts.

The observed physical phenomena such as large entropy per baryon and remarkable degree of isotropy of the cosmic background radiation, suggest dissipative effects in cosmology. It has been argued for a long time that the dissipation process in the early stage of cosmic expansion may well account for the high degree of isotropy, we observe today. Dissipative effects including both bulk and shear viscosities play a very significant role to study the early evolution of universe.~\citet{rf311} developed the first relativistic theory of non-equilibrium thermodynamics to study the effect of viscosity.~\citet{rf312} investigated that presence of bulk viscosity leads to inflationary like solutions in general relativity. Another peculiar characteristic of bulk viscosity is that it acts like a negative energy field in an expanding universe~\citep{rf313}. The effect of bulk viscosity  on  the  cosmological  evolution  has  been  investigated by number of authors viz.~\citet{rf314,rf315,rf316,rf317,rf318,rf319}.

The cosmological constant problem is one of the outstanding problem in cosmology~\citep{rf320}. It has a two fold meaning: It is a problem of fundamental physics because the value of cosmological constant is tied to a vacuum energy density, on the other hand, the cosmological constant tells us about the large scale behavior of the universe, since a small cosmological constant implies that the observable universe is a big one and nearly flat. The problem is that there is an enormous discrepancy predicted by quantum field theory of standard model and the cosmological observed value of $\Lambda$~\citep{rf321}. Its value is considered to be the order of 10$^{-58}$cm$^{-2}$~\citep{rf322}.~\citet{rf323} has suggested that $\Lambda$ is a function of temperature and is related to spontaneous symmetry breaking process. Hence $\Lambda$ is considered to be the function of time for spatially homogeneous expanding universe.~\citet{rf324} have shown that the cosmological relativistic theory predicts    $\Lambda$ = $1.934\times10^{-35}$s$^{-2}$ which is in agreement with the measurements recently obtained by the High-Z supernova team and supernova cosmological project ~\citep{rf325}. A number of authors have studied cosmological models with time dependent cosmological term viz.~\citet{rf326,rf327,rf328,rf329,rf319,rf330,rf331,rf332,rf333}.

In this paper, we investigate Bianchi type V cosmological model for viscous fluid  distribution   with varying cosmological term $\Lambda$. We examine a cosmological scenario proposing a variation law for Hubble parameter $H$ in the background of homogeneous, anisotropic Bianchi type V space-time. The model isotropizes asymptotically and the presence of shear viscosity accelerates the isotropization. The model describes a unified expansion history of the universe indicating decelerating and accelerating phase both.

\section{Metric and Field Equations}
We consider Bianchi type V space-time in orthogonal form represented by the line-element
\be\label{eq301}
ds^2=-dt^2+A^2(t)dx^2+e^{2\alpha x}\left\{B^2(t)dy^2+C^2(t)dz^2\right\},
\ee
where $\alpha$ is a constant. Cosmic matter is assumed to be a viscous fluid given by the energy-momentum tensor
\be\label{eq302}
T_{i}^{j}=(\rho+\bar{p})v_iv^j+\bar{p}g_{i}^{j}-2\eta\sigma_{i}^{j},
\ee
where $\bar{p}$ is the effective pressure given by
\be\label{eq303}
\bar{p}=p-\zeta {v^i}_{;i},
\ee
satisfying linear equation of state
\be\label{eq304}
p=\omega\rho,~~~0\leq\omega\leq1.
\ee
$\rho$ being matter energy density, p the isotropic pressure, $\eta$ and $\zeta$ are coefficients of shear and bulk viscosities respectively, $v^{i}$ the flow vector of the fluid satisfying $v_iv^i=-1$ and $\sigma_{ij}$ is shear tensor given by
\be\label{eq305}
\sigma_{ij}=\frac{1}{2}\left(v_{i;j}+v_{j;i}\right)+\frac{1}{2}\left(\dot{v}_{i}v_{j}+v_{i}\dot{v}_{j}\right)-\frac{1}{3}{v^k}_{;k}\left(g_{ij}+v_iv_j\right).
\ee
where $\dot{v}_i=v_{i;j}v^j$. We assume the coordinates to be comoving so that $v^1=0=v^2=v^3$, $v^4=1$. The Einstein's field equations (in gravitational units $8\pi G = c = 1$) with time varying cosmological term $\Lambda(t)$ are given by
\be\label{eq306}
R_{ij}-\frac{1}{2}R_k^kg_{ij}=-T_{ij}+\Lambda(t) g_{ij}.
\ee
The field equations (\ref{eq306}) for the line element (\ref{eq301}) give rise to
\be\label{eq307}
p-\left(\zeta-\frac{2}{3}\eta\right)\theta-\Lambda=\frac{\alpha^2}{A^2}-\frac{\ddot{B}}B-\frac{\ddot{C}}C-\frac{\dot{B}\dot{C}}{BC}+2\eta\frac{\dot{A}}{A},\\
\ee
\be\label{eq308}
p-\left(\zeta-\frac{2}{3}\eta\right)\theta-\Lambda=\frac{\alpha^2}{A^2}-\frac{\ddot{C}}C-\frac{\ddot{A}}A-\frac{\dot{C}\dot{A}}{CA}+2\eta\frac{\dot{B}}{B},\\
\ee
\be\label{eq309}
p-\left(\zeta-\frac{2}{3}\eta\right)\theta-\Lambda=\frac{\alpha^2}{A^2}-\frac{\ddot{A}}A-\frac{\ddot{B}}B-\frac{\dot{A}\dot{B}}{AB}+2\eta\frac{\dot{C}}{C},\\
\ee
\be\label{eq310}
\rho+\Lambda=-\frac{3\alpha^2}{A^2}+\frac{\dot{A}\dot{B}}{AB}+\frac{\dot{B}\dot{C}}{BC}+\frac{\dot{A}\dot{C}}{AC},\\
\ee
\be\label{eq311}
\frac{2\dot{A}}{A}=\frac{\dot{B}}{B}+\frac{\dot{C}}{C}.
\ee
where overhead (.) denotes ordinary time derivative and $\theta={v^i}_{;i}$ is volume expansion scalar.
Since covariant divergence of Einstein tensor $G_{ij}=R_{ij}-\frac{1}{2}R_k^kg_{ij}$ vanishes identically, we obtain
\be\label{eq312}
\dot{\rho}+(\rho+\bar{p})\left(\frac{\dot{A}}{A}+\frac{\dot{B}}{B}+\frac{\dot{C}}{C}\right)+\dot{\Lambda}=4\eta\sigma^2
\ee
where $\sigma$ is shear scalar given by
\be\label{eq313}
\sigma^2=\frac{1}{2}\sigma_{ij}\sigma^{ij},
\ee
We define average scale factor $R$ for Bianchi type V space-time as
\be\label{eq314}
R^3=ABC.
\ee
Generalized Hubble parameter $H$ and generalized deceleration parameter $q$ are defined as
\be\label{eq315}
H=\frac{\dot{R}}{R}=\frac{1}{3}(H_1+H_2+H_3)
\ee
and
\be\label{eq316}
q=-\frac{R\ddot{R}}{\dot{R}^2}=-\frac{\dot{H}}{H^2}-1,
\ee
where $H_1=\frac{\dot{A}}{A}$, $H_2=\frac{\dot{B}}{B}$ and $H_3=\frac{\dot{C}}{C}$ are directional Hubble's factors along $x$, $y$ and $z$ directions respectively.

Expansion scalar $\theta$ and shear tensor $\sigma_i^j$ for the metric (\ref{eq301}) lead to
\be\label{eq317}
\theta=3H,
\ee
\be\label{eq318}
\sigma_1^1=H_1-H,~\sigma_2^2=H_2-H,~\sigma_3^3=H_3-H,~\sigma_4^4=0.
\ee
Shear scalar $\sigma$ is given by
\bea
\sigma^2&=&\frac{1}{2}({H_1}^2+{H_2}^2+{H_3}^2)-\frac{3}{2}H^2 \notag\\
\label{eq319}
&=&\frac{1}{6}\left\{\left(\frac{\dot{A}}{A}-\frac{\dot{B}}{B}\right)^2+\left(\frac{\dot{B}}{B}-\frac{\dot{C}}{C}\right)^2+\left(\frac{\dot{C}}{C}-\frac{\dot{A}}{A}\right)^2\right\}.
\eea

From equations (\ref{eq307})-(\ref{eq309}), we have
\be\label{eq322}
\frac{\ddot{A}}{A}-\frac{\ddot{B}}{B}+\left(\frac{\dot{A}}{A}-\frac{\dot{B}}{B}\right)\left(\frac{\dot{C}}{C}+2\eta\right)=0
\ee
\be\label{eq323}
\frac{\ddot{B}}{B}-\frac{\ddot{C}}{C}+\left(\frac{\dot{B}}{B}-\frac{\dot{C}}{C}\right)\left(\frac{\dot{A}}{A}+2\eta\right)=0
\ee
Equations (\ref{eq322}) and (\ref{eq323}) with the help of equation (\ref{eq311}) reduce to a single equation
\be\label{eq324}
\frac{\ddot{B}}{B}-\frac{\ddot{C}}{C}+\left(\frac{\dot{B}}{B}-\frac{\dot{C}}{C}\right)\left\{\frac{1}{2}\left(\frac{\dot{B}}{B}+\frac{\dot{C}}{C}\right)+2\eta\right\}=0
\ee
which on integration leads to
\be\label{eq325}
\frac{\dot{B}}{B}-\frac{\dot{C}}{C}=\frac{2k}{R^3}e^{-2\int\eta~dt},
\ee
where $k=$ constant. From equations (\ref{eq311}), (\ref{eq317}) and (\ref{eq325}), we get
\be\label{eq326}
\frac{\dot{A}}{A}=\frac{\dot{R}}{R},~~
\frac{\dot{B}}{B}=\frac{\dot{R}}{R}+\frac{k}{R^3}e^{-2\int\eta~dt},~~
\frac{\dot{C}}{C}=\frac{\dot{R}}{R}-\frac{k}{R^3}e^{-2\int\eta~dt}.
\ee
We propose the form of shear and bulk viscosity as considered by~\citet{rf336,rf337}
\be\label{eq327}
\eta=3\eta_0\frac{\dot{R}}{R}~\mbox{and}~\zeta=\zeta_0+\zeta_1\frac{\dot{R}}{R}+\zeta_2\frac{\ddot{R}}{R},
\ee
where $\eta_0,\zeta_0,\zeta_1,\zeta_2$ are constants. For this choice, equation (\ref{eq326}) reduces to
\be\label{eq328}
\frac{\dot{A}}{A}=\frac{\dot{R}}{R},~~
\frac{\dot{B}}{B}=\frac{\dot{R}}{R}+\frac{k}{R^{3+6\eta_0}},~~
\frac{\dot{C}}{C}=\frac{\dot{R}}{R}-\frac{k}{R^{3+6\eta_0}}.
\ee
In this case, we obtain
\be\label{eq329}
\sigma_1^1=0,~~\sigma_2^2=\frac{k}{R^{3+6\eta_0}},~~\sigma_3^3=-\frac{k}{R^{3+6\eta_0}}
\ee
and
\be\label{eq330}
\sigma=\frac{k}{R^{3+6\eta_0}}
\ee
We observe that the presence of shear viscosity accelerates the isotropization process.\\
Equations (\ref{eq307})-(\ref{eq310}) and (\ref{eq312}) can be written in terms of $H$, $\sigma$ and $q$ as
\be\label{eq331}
\bar{p}-\Lambda=(2q-1)H^2-\sigma^2+\frac{\alpha^2}{R^2},
\ee
\be\label{eq332}
\rho+\Lambda=3H^2-\sigma^2-\frac{3\alpha^2}{R^2},
\ee
\be\label{eq333}
\dot{\rho}+3(\rho+\bar{p})H+\dot{\Lambda}=12\eta_0H\sigma^2.
\ee
\section{Solution of the Field Equations}
We observe that equation (\ref{eq325}) is a single equation involving two unknowns B and C. We require one more condition to close the system. We assume a variation law for the Hubble parameter~\citep{ex1} as
\be\label{eq334}
H(R)=a(R^{-n}+1),~a>0,~n>1~\mbox{being constants}.
\ee
It gives a model of the universe describing decelerating expansion followed by late time acceleration consistent with observations. For this choice, the deceleration parameter $q$ comes out to be
\be\label{eq335}
q=\frac{n}{R^n+1}-1
\ee
We assume that $R = 0$ for $t = 0$. We observe that for $R \approx 0$, $q \approx n - 1 > 0$, $q = 0 $ for $R^{n}= n -1$ and for $R^{n}> n - 1$, $q < 0$.

Thus, we have a model of universe which begins with a decelerating expansion and evolves into a late time accelerating universe which is in agreement with S Ne Ia astronomical observations~\citep{rf338}.

Using (\ref{eq330}), (\ref{eq334}) and (\ref{eq335}) in (\ref{eq331}) and (\ref{eq332}), we get
\be
p=\frac{\alpha^2}{R^2}+a^2(R^{-n}+1)^2\left(\frac{2n}{R^n+1}-3\right)-\frac{k^2}{R^{6+12\eta_0}}+3\zeta H+\Lambda
\ee
i.e.
\bea
p&=&3\left\{\zeta_0a(R^{-n}+1)+\zeta_1a^2(R^{-n}+1)^2-\zeta_2a^3(R^{-n}+1)^3\left(1-\frac{n}{R^n+1}\right)\right\}\notag\\
\label{eq337}
&&+\frac{\alpha^2}{R^2}+a^2(R^{-n}+1)^2\left(\frac{2n}{R^n+1}-3\right)-\frac{k^2}{R^{6+12\eta_0}}+\Lambda
\eea
and
\be\label{eq338}
\rho=-\frac{3\alpha^2}{R^2}+3a^2(R^{-n}+1)^2-\frac{k^2}{R^{6+12\eta_0}}-\Lambda
\ee
We observe that the model has singularity at $t = 0$ (i.e. $R = 0$). For large values of $t$ (i.e. $R\rightarrow\infty$), we get
\be\label{eq339}
\rho=3a^2-\Lambda
\ee
and
\be\label{eq340}
p=-3a^2+3(\zeta_0a+\zeta_1a^2+\zeta_2a^3)+\Lambda.
\ee
From (\ref{eq330}), we obtain
\be\label{eq341}
\dot{\sigma}=-(3+6\eta_0)\sigma H.
\ee
Thus the energy density associated with anisotropy $\sigma$ decays due to expansion by converting into photons and presence of shear viscosity accelerates this decay. From equations (\ref{eq331}) and (\ref{eq332}), we get
\be\label{eq342}
\frac{\ddot{R}}{R}=-\frac{1}{6}(\rho+3p)-\frac{2\sigma^2}{3}+\frac{\zeta\theta}{3}+\frac{\Lambda}{3}.
\ee
We observe that bulk viscosity and positive $\Lambda$ contribute positively in driving acceleration of the universe whereas active gravitational mass density and anisotropy arrest this acceleration.

Equation (\ref{eq334}) after integration leads to
\be\label{eq343}
R^n=e^{na(t+t_1)}-1
\ee
where $t_{1}$ is constant of integration. Assuming $R = 0$ for $t = 0$, we get $t_{1} = 0$ and
\be\label{eq344}
R^n=e^{nat}-1
\ee
For the model, the matter density $\rho$ and cosmological term $\Lambda$ are given by
\bea
\rho&=&\frac{1}{1+\omega}\left\{\frac{2na^2e^{nat}}{(e^{nat}-1)^2}-\frac{2k^2}{(e^{nat}-1)^{\frac{6+12\eta_0}{n}}}-\frac{2\alpha^2}{(e^{nat}-1)^\frac{2}{n}}+\frac{3\zeta_0ae^{nat}}{(e^{nat}-1)}\right\}\notag\\
\label{eq345}
&&+\frac{1}{1+\omega}\left\{\frac{3\zeta_1a^2e^{2nat}}{(e^{nat}-1)^2}+\frac{3\zeta_2a^3e^{2nat}(e^{nat}-n)}{(e^{nat}-1)^3}\right\}
\eea
\bea
\Lambda&=&\frac{a^2e^{nat}}{(e^{nat}-1)^2}\left(3e^{nat}-\frac{2n}{\omega+1}\right)+\frac{1}{(1+\omega)}\left\{\frac{(1-\omega)k^2}{(e^{nat}-1)^{\frac{6+12\eta_0}{n}}}-\frac{(3\omega+1)\alpha^2}{(e^{nat}-1)^{\frac{2}{n}}}\right\}\notag\\
\label{eq346}
&&-\frac{3}{1+\omega}\left\{\frac{\zeta_0a}{1-e^{-nat}}+\frac{\zeta_1a^2}{(1-e^{-nat})^2}+\frac{\zeta_2a^3(1-ne^{-nat})}{(1-e^{-nat})^3}\right\}
\eea
Expansion $\theta$, shear $\sigma$, deceleration parameter $q$, coefficient of shear viscosity $\eta$ and bulk viscosity $\zeta$ of the model take the form
\be\label{eq347}
\theta=\frac{3a}{1-e^{-nat}}
\ee
\be\label{eq348}
\sigma=\frac{k}{(e^{nat}-1)^{\frac{3+6\eta_0}{n}}}
\ee
\be\label{eq349}
q=\frac{n}{e^{nat}}-1
\ee
\be\label{eq350}
\eta=\frac{3\eta_0a}{1-e^{-nat}}
\ee
\be\label{eq351}
\zeta=\zeta_0+\frac{\zeta_1a}{1-e^{-nat}}+\frac{\zeta_2a^2(1-ne^{-nat})}{(1-e^{-nat})^2}.
\ee
\section{Discussion}
\begin{figure}
\centering
\includegraphics[width=4in]{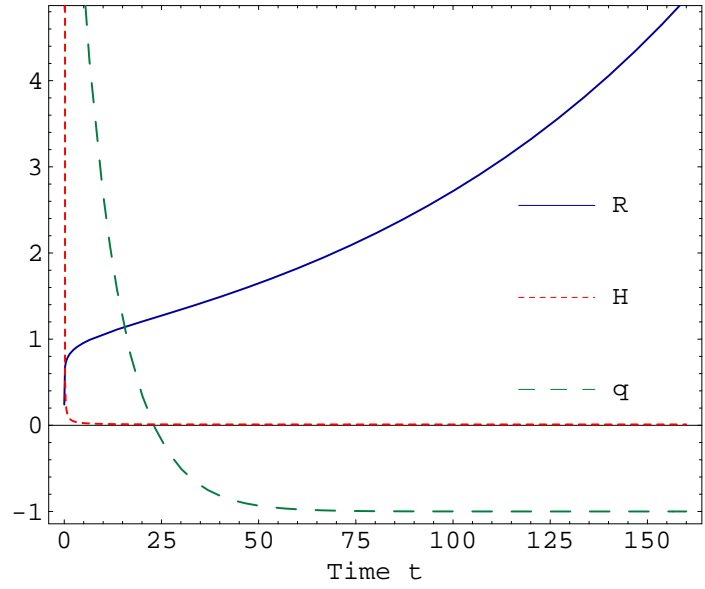}\\
 \caption{ Variation of Scale Factor $R$, Hubble Parameter $H$ and Deceleration Parameter $q$ with cosmic time $t$.}
\label{fig1}
\end{figure}

\begin{figure}
\centering
\includegraphics[width=4in]{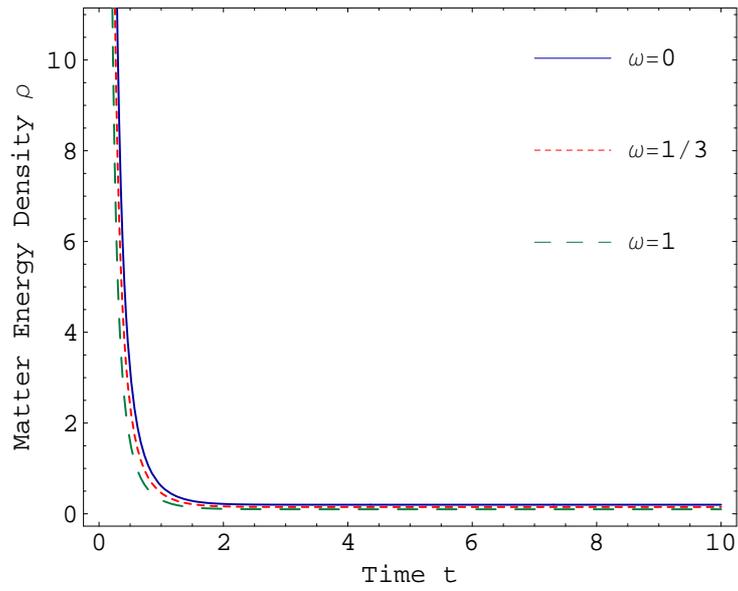}\\
 \caption{Variation of Matter Energy Density $\rho$ with cosmic time $t$.}
\label{fig2}
\end{figure}

\begin{figure}
\centering
\includegraphics[width=4in]{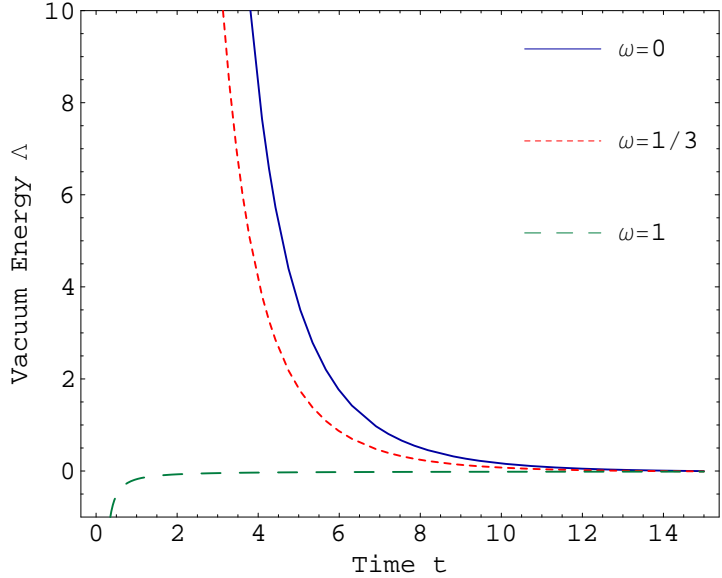}\\
 \caption{Variation of Vacuum Energy $\Lambda$ with cosmic time $t$.}
\label{fig3}
\end{figure}

We observe that matter density $\rho$, expansion $\theta$, shear $\sigma$, cosmological constant $\Lambda$, coefficients of shear and bulk viscosities all diverge at $t = 0$. The model starts with a big-bang from  its  singular  state  at $t = 0$ and continues to expand till $t = \infty$. For $t \approx 0$, deceleration parameter $q = (n-1) > 0$. Therefore, the model starts with a decelerating expansion and after a lapse of finite time $t_q=\frac{\ln n}{na}$, decelerating phase in the model comes to an end. For $t > t_{q}$, $q < 0$ so that the model enters accelerating regime of expansion. In the limit of large times i.e. $t \rightarrow\infty$, $\rho\rightarrow\frac{3}{1+\omega}(\zeta_0a+\zeta_1a^2+\zeta_2a^3)$, $\Lambda\rightarrow3a^2-\frac{3}{1+\omega}(\zeta_0a+\zeta_1a^2+\zeta_2a^3)$, $H\rightarrow a$, $\sigma\rightarrow0$ and $q\rightarrow-1$. We observe that the presence of bulk viscosity prevents the model to tend to a de-Sitter universe and the matter density to become negligible asymptotically. Coefficients of shear and bulk viscosities tend to genuine constants for large values of $t$. For the model
\be\label{eq352}
\frac{\sigma}{\theta}=\frac{k(1-e^{-nat})}{3a(e^{nat}-1)^{\frac{3+6\eta_0}{n}}}
\ee
For large values of $t$,   $\frac{\sigma}{\theta}\rightarrow0$ implying that the model approaches isotropy at late times. We observe that the presence of shear viscosity accelerates the process of isotropization. For illustrative purposes, the time variation of different cosmological parameters are shown graphically in Figs. \ref{fig1}, \ref{fig2} and \ref{fig3}. We obtain the present value of the cosmological term $\Lambda$ for the age of the universe $t_0=13.69$ Gyr~\citep{ex2} as
$$
\Lambda\approx 0.0125~\mbox{Gyr$^{-2}$}~\approx1.252\times 10^{-35}~\mbox{S$^{-2}$}
$$
which is in agreement with observational values.
\section{Conclusion}
In this paper, we examine a cosmological scenario proposing a variation law for Hubble parameter $H$ in the background of homogeneous, anisotropic Bianchi type V space-time which contains viscous fluid matter distribution. The model starts with a big bang from its singular state with decelerating expansion and after a lapse of finite time, expansion in the model changes from decelerating phase to accelerating one. Because of bulk viscosity, matter density does not become negligible and the model does not tend to a de-Sitter universe for large values of $t$. The model isotropizes asymptotically and the presence of shear viscosity accelerates the isotropization. The model describes a unified expansion history of the universe which starts with decelerating expansion and the expansion accelerates at late time. Decelerating expansion at the initial epoch provides obvious provision for the formation of large structures in the universe. The formation of structures in the universe is better supported by decelerating expansion. Thus the resulting model is astrophysically relevant. Also late time acceleration is in agreement with the observations of 16 type Ia supernovae made by Hubble Space Telescope (HST)~\citep{rf338}. The agreement with the observed universe is just qualitative. Precise observational tests are required to verify or disprove the model.
%

\end{document}